\documentclass[%
reprint,
 amsmath,amssymb,
 aps,
]{revtex4-2}

\usepackage{graphicx}
\usepackage{dcolumn}
\usepackage{bm}
\usepackage{multirow}
\usepackage{caption}
\usepackage{blindtext}
\usepackage{appendix}
\usepackage{graphicx}
\usepackage{amsmath}
\usepackage{xcolor}

\newcommand{\bbt}{\textcolor{black}}
\begin{document}
\title{K-shell ionization cross sections of Cu, Zn and Ge by 3-5 MeV/U Si-ion bombardment }

\author{Shashank Singh$^1$, Mumtaz Oswal$^2$, Sunil Kumar$^3$, K.P. Singh$^1$, D. Mitra$^4$  and T. Nandi$^{5*}$}
\affiliation{$^1$Department of Physics, Panjab University, Chandigarh-160014, India.}
\affiliation{$^2$Department of Physics, Dev Samaj College for Women, Sector 45 B, Chandigarh-160014, India.}
\affiliation{$^3$Govt. Degree college, Banjar, Kullu, Himachal Pradesh-175123, India.}
\affiliation{$^4$Department of Physics, University of Kalyani, Kalyani, Nadia-741235, West Bengal, India. }
\affiliation{$^{5}$Inter-University Accelerator Centre, Aruna Asaf Ali Marg, Near Vasant Kunj, New Delhi-110067, India.}
\thanks {Email:\hspace{0.0cm} nanditapan@gmail.com. Present address: 1003 Regal, Mapsko Royal Ville, Sector-82, Gurgaon-122004, India.}

\begin{abstract}
The K x-ray spectra of different targets (Cu, Zn, and Ge) induced by \bbt{3 to 5 MeV/u} Si projectile ions have been measured to determine the K-shell ionization cross-section. A significant difference is observed between the measurements and theoretical estimates, \bbt{where the} theoretical ones are about \textcolor{black}{28-35\%} of the experimental results. Such difference is reduced to a good extent 51-56\% if multiple ionization effects are taken into account. Remaining discrepancy may be attributed to the electron capture contribution. 
\end{abstract}
\maketitle
\section{Introduction}
\indent Ionization dynamics of target atoms by energetic heavy ions is important in several fields of research such as material analysis, material engineering, atomic and nuclear physics, accelerator physics, \bbt{biophysics}, medical science, etc. The precise data of ionization cross-section of target atoms are required in case of heavy-ion application in \bbt{particle-induced} x-ray emission (PIXE) \cite{miranda2007x} and in heavy-ion \bbt{tumor} therapy \cite{kraft2000tumor}. Appropriate knowledge of K-shell ionization \bbt{is essential} to determine the elemental concentration during PIXE analysis and to estimate the direct damage of the \bbt{tumor} by the projectiles. Besides the target ionization, the effect of secondary electrons during heavy-ion impact in the patient's body is very significant \bbt{leading} to much greater damage than the direct damage by the incident ions. The secondary electron yield is found to be proportional to the rate of energy loss of the incident particles \cite{sternglass1957theory}. This energy loss is connected to the excitation and ionization processes of the target atoms and projectile ions, where inner shell ionization is most vital. \\
\indent Inner shell ionization of target atoms are being carried out for a long time with light as well as heavy projectiles, for examples
\cite{benka1978tables,orlic1994experimental,kadhane2003k,lapicki2005status,zhou2013k,msimanga2016k,kumar2017shell,oswal2018x,hazim2020high,oswal2020experimental,miranda2020total}. 
\textcolor{black}{It has enabled the research community to study the processes} like ionization, excitation, multiple ionization \cite{lapicki2004effects}, radiative decay, Auger-decay \cite{dahl1976auger}, changes in atomic parameters, intrashell coupling effect (in L- and M-shell not in K-shell) \cite{pajek2003multiple,sarkadi1981possible}, etc at different energy regimes. In the present work, we have measured the K-shell x-ray yields of target atoms in three projectile-target systems, i.e., Si + Cu, Si + Zn, and Si + Ge. Using K x-ray yields, we have determined the K-shell ionization cross-section of the target atoms. It is observed that the present measurements are about a factor of three higher than the theoretical direct ionization cross-sections. Such a large discrepancy is a matter of concern to date and needs to be resolved not only for fundamental understanding, but also for practical implications in  heavy-ion induced x-ray emission techniques for elemental analysis.
\section{Experimental details}
The experiment was performed in the atomic physics beam line of 15 UD Pelletron which is situated at Inter-University Accelerator Centre, New Delhi (India). Si ion beam of charge state 8$^+$ for beam energies 84, 90, 98, 107 MeV and charge state 12$^+$ for beam energies 118, 128, 140 MeV was obtained from Pelletron to bombard the natural Cu, Zn, and Ge targets. The vacuum of the order of $10^{-6}$ Torr was maintained in the chamber using a turbo-molecular pump. Two silicon surface barrier detectors were placed at $\pm7.5^\circ$ with respect to beam direction to normalize the charge. A Si(Li) solid state detector was placed outside the chamber at $125^\circ$ with respect to beam direction and distance of 170 mm from the target. A collimator of 5 mm diameter was placed in front of the detector inside the chamber. The thickness of the Mylar window of the chamber for the detector was 6 $\mu m$. The specification of the detector (ORTEC, Oak Ridge, Tennessee, USA) is as follows: thickness 5 mm, diameter 10 mm, the thickness of Be window 25 $\mu m$ and energy resolution ~200 eV for Mn $K_\alpha$ x-rays.
\begin{figure*}
\centering
\includegraphics[width=17.5cm,height=8cm]{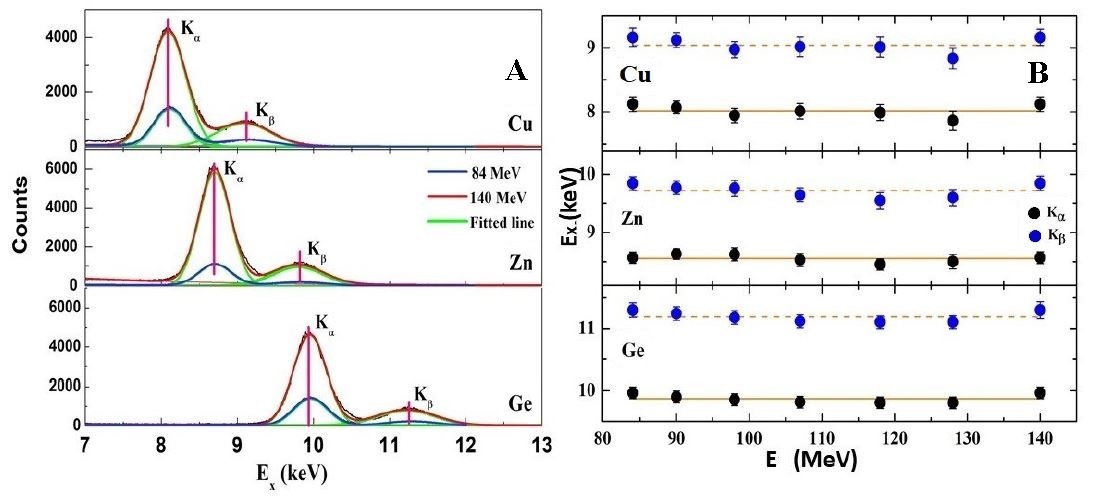}
\caption{(A) Typical K x-ray spectra of natural Cu, Zn, and Ge targets when bombarded with 84 and 140 MeV $^{28}$Si ions; (B) x-ray peak energy shift in these targets as a function of the ion-beam energies of $^{28}$Si ions. The solid and dotted orange lines in the right panel show the mean value of the $K_\alpha$ and $K_\beta$ peak energies.}
\label{Spectra}
\end{figure*}
\indent The energy calibration of the detector was done before and after the experiment using the $^{55}$Fe, $^{57}$Co and $^{241}$Am radioactive sources. \textcolor{black}{The target surface was placed at $90^{\circ}$ to the beam direction (normal to the target surface is collinear to the beam direction)} on a rectangular steel ladder which could move horizontal and vertical direction with the help of a stepper motor. The spectroscopically pure (99.999\%) thin targets of natural Cu, Zn, and Ge were made on the carbon backing using vacuum deposition technique. The thickness of Cu, Zn, Ge, and carbon backing was 25 $\mu g/cm^2$, 14.4 $\mu g/cm^2$, 99 $\mu g/cm^2$, and 20 $\mu g/cm^2$, respectively. The thicknesses of targets were measured using the energy loss method using $^{241}$Am radioactive source. The data was acquired using a PC based software developed at IUAC \cite{subramaniam2010data}. The beam current was kept below 1 nA to avoid pile-up effects and damage to the targets. A semi-empirical fitted relative efficiency curve used for the present measurement can be seen in Oswal et.al. \cite{oswal2020experimental}.
\section{DATA ANALYSIS, RESULT, AND DISCUSSION}
\indent Typical K x-ray spectra of Cu, Zn, and Ge bombarded with 84 MeV and 140 MeV Si ions are shown in Fig.\ref{Spectra}. The spectra were analyzed with a nonlinear least-squares fitting method considering a Gaussian line shape for the x-ray peaks and a linear background fitting. The x-ray production cross-sections for the K x-ray lines were determined from the relation,
\begin{equation}
\sigma_i^x=\frac{Y_i^x A}{N_A\epsilon n_p t \beta} \label{eqn:3}
\end{equation}
\noindent here $Y_i^x$ is the intensity of $i^{th}$ x-ray peak ($i$ = K$_\alpha$, K$_\beta$). $A$ is the atomic weight of the target. $N_A$ and $n_p$ denote the Avogadro number and the number of incident projectiles, respectively. $\epsilon$, $t$, and $\beta$ represent the effective efficiency of the x-ray detector, the target thickness in $\mu g/cm^2$, and the correction factor for energy-loss of the incident projectile and absorption of emitted x-rays in the target element, respectively.
The sum of $\sigma_{K_\alpha}^x$ and $\sigma_{K_\beta}^x$ gives a measure of the total K x-ray production cross-section as given in Table \ref{EXPERIMENTAL CROSS-SECTION}.\\
\begin{figure*}
\centering
\includegraphics[width=16.0cm,height=6.0cm]{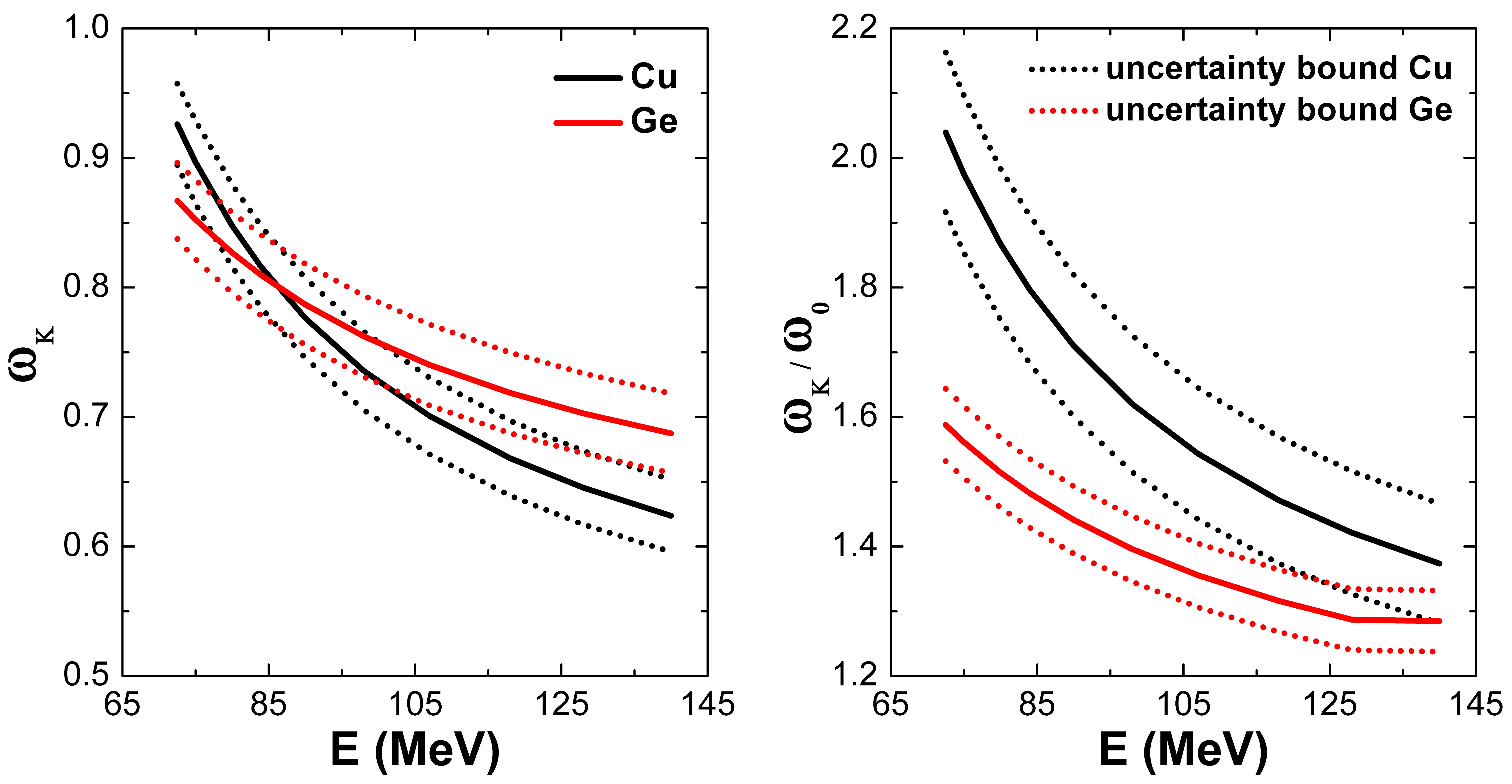}
\caption{Variation of $\omega_K$ and ratio of $\omega_K$ to $\omega_0$ with respect to projectile energy (E)}
\label{wk_vs_E}
\end{figure*}
\indent It is now well known that heavy ions produce simultaneous multiple ionization (SMI) in several shells while travelling through the target. SMI of L-shells along with a vacancy in K-shell will influence the value of K-shell fluorescence yield ($\omega_K$) to a considerable extent. Instead of using rigorous Hartree-Fock-Slater calculation for the $K_\alpha$ and $K_\beta$ peak shift due to the SMI effect, we employ a simple model of \textcite{burch1974simple}. According to it, the energy shift of $K_\alpha$ and $K_\beta$ lines per $2p$ vacancy with respect to corresponding diagram lines are 1.66$Z_L$ and 4.18$Z_L$ eV, respectively, where $Z_L$ = $Z_2$ -- 4.15; $Z_2$ is the atomic  number of the target element. It is clear from Fig.\ref{Spectra}(A) that $K_\alpha$ and $K_\beta$ lines are well resolved for all the targets used in the present measurements.\\
\indent In order to visualize the centroid shift due to the SMI effect, we have plotted the $K_\alpha$ and $K_\beta$ energies versus beam energy for all the targets in Fig.\ref{Spectra}(B). These data are also given in Table \ref{AVERAGE PEAK ENERGY}. We notice, for all the targets, the corresponding centroid energies do not vary much with the beam energies used. The average K$_\alpha$ peak energies of $Cu$, $Zn$, and $Ge$ are 8.01 $\pm$ 0.1, 8.55 $\pm$ 0.1, and 9.87 $\pm$ 0.1 keV, respectively and are close to the corresponding diagram K$_\alpha$ lines at 8.03, 8.62, 9.86 keV. \textcolor{black}{In contrast, this picture for K$_\beta$ lines is rather distinctive.} The mean of the measured K$_\beta$ lines 9.04 $\pm$ 0.13, 9.72 $\pm$ 0.12, and 11.2 $\pm$ 0.1 keV for Cu, Zn and Ge, respectively are higher than the corresponding diagram K$_\beta$ lines at 8.905, 9.572, and 10.982 keV. Thus, the difference between the measured $K_\beta$ and the diagram $K_\beta$ lines for Cu, Zn, and Ge are 135, 148, and 218 eV, respectively. These values are somewhat larger than the energy shift per $2p$ vacancy for $K_\beta$ lines, which are 104, 108, 116 eV, respectively, for Cu, Zn, and Ge. \\
\indent Energy shifts along with the measurement uncertainty mentioned above imply that on average two vacancies occur in $2p$ shells during the present collisions. This picture corroborates well the scenario in the $K_\alpha$ case too as the energy shift per $2p$ vacancy in Cu, Zn and $Ge$ for $K_\alpha$ line are only 41, 43, 46 eV, respectively, and the energy shift due to two $2p$ vacancies will be smeared in its measurement uncertainty of about 100 eV. Thus the SMI must be included in data analysis.
\begin{figure*}
\centering
\includegraphics[width=18.0cm,height=6.0cm]{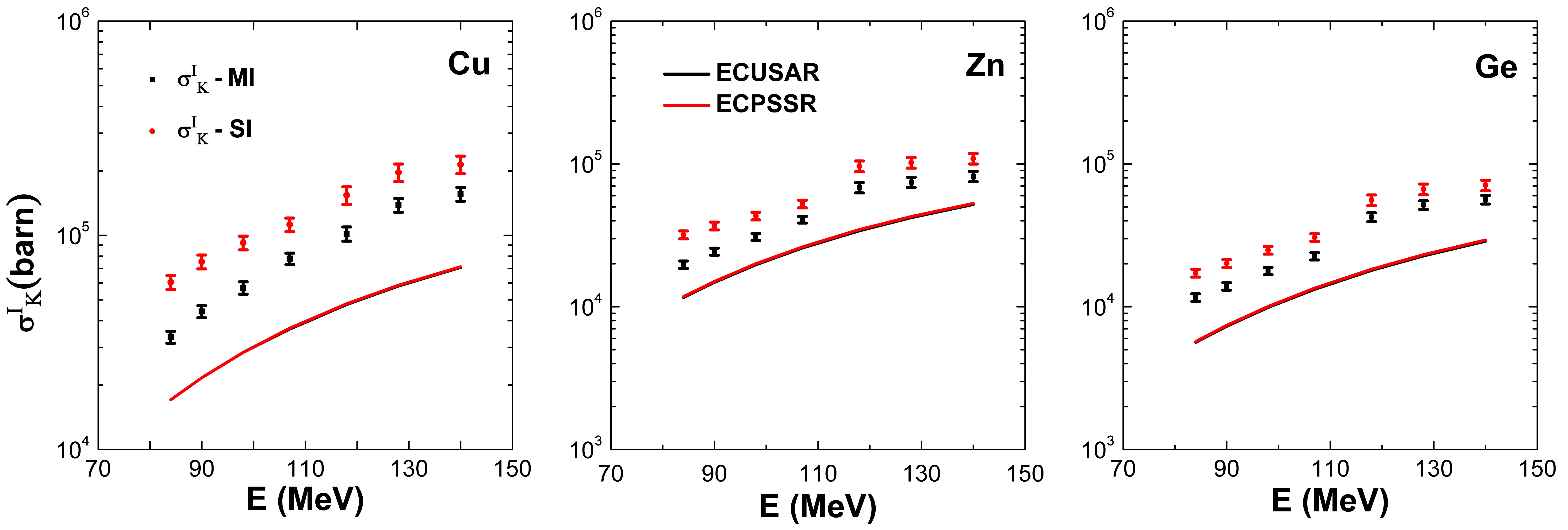}
\caption{Comparison of experimental K shell ionization cross-sections ($\sigma_K$) of different targets bombarded by the $^{28}{Si}$ ions as a function of ion-beam energies with the direct ionization cross-sections from ECUSAR \cite{lapicki2004effects} and ECPSSR \cite{brandt1981energy}. Here $\sigma_K$-SI and $\sigma_K$-MI indicate K shell ionization cross-sections with a single K-vacancy and with a single K-vacancy along with several vacancies in higher shells, respectively. } 
\label{Expt and DI theories}
\end{figure*}
%
%
\par Theoretically, K x-ray production cross-section ($\sigma_K^x$) can be obtained using the relation \cite{kadhane2003k}
\begin{equation}
\sigma^x_{K}= \omega_K\sigma^I_{K} \label{eqn:Sig-KI}
\end{equation}
here $\sigma^I_{K}$ is K-shell ionization cross-section, $\omega_K$ is the K shell fluorescence yield in the presence of SMI effect in L-shell. Single vacancy fluorescence yield $\omega_K^0$ given by Krause \cite{krause1979atomic} has been used. Hence, to extract the K shell ionization cross-section from the measured x-ray production cross-section one needs the accurate knowledge of $\omega_K$. \\
\indent To estimate $\omega_K$ amidst the SMI process discussed above, we follow the description of Lapicki $et. al.$ \cite{lapicki2004effects} using an assumption that each electron in a manifold of outer subshells is ionized with an identical probability $P$ and correct $\omega_K$ in presence of SMI process becomes
\begin{equation}
 \omega_K=\frac{\omega_K^0}{1-P(1-\omega_K^0)}.
\end{equation}
With 
\begin{equation}
 P= q_m^2(1-\frac{0.225}{v_1^2})\times\frac{1}{1.8v_1^2}
\end{equation}
\noindent here $v_1=6.351 [E/A_1]^{1/2}$ (E and $A_1$ are projectile energy and mass in MeV and amu units, respectively) is the projectile velocity. $q_m$ is the mean charge state of the projectile ion inside the target.\\
\textcolor{black}{\indent The uncertainty in $\omega_K$ can be estimated from the following expression}
\begin{multline}
\frac{\Delta\omega_K}{\omega_K}=\frac{\Delta\omega_K^0}{\omega_K^0}+\frac{P}{[1-P(1-\omega_K^0)]}\\
\times[\frac{\Delta P}{p}\times (1-\omega_K^0)-\frac{\Delta w_k^0}{\omega_K^0}]
\end{multline}
where
\begin{equation}
\begin{aligned}
\frac{\Delta P}{P}=\frac{2 \Delta q_1}{q_1}.
\end{aligned}
\end{equation}
and the uncertainty in $\frac{\omega_K}{\omega^0_K}$ can be obtained as follows
\begin{equation}
\begin{aligned}
\frac{{\Delta (\omega_K/\omega^0_K)}}{\omega_K/\omega^0_K} = \frac{\Delta\omega_K}{\omega_K}[1-\{\frac{\omega^0_K}{\omega_K}\}^2]
\end{aligned}
\end{equation}
The projectile velocity $v_1$ can be defined very precisely and thus its uncertainty is nominal ($< 1\%$) and taken as just a constant here. For Cu, $\frac{\Delta \omega_K^0}{\omega_K^0}$ is $\approx 5\%$ and this is $\approx 3\%$ for Zn and Ge.
If we assume $\frac{\Delta q_1}{q_1} is \approx 3\%$ (its estimation and probable uncertainty will be discussed later), $\frac{\Delta \omega_K}{\omega_K}$ turns out to be $\approx 6\%$. Fig. \ref{wk_vs_E} shows the variation of $\omega_K$ and $\frac{\omega_K}{\omega_K^0}$ as a function of projectile energy (E). The uncertainty bound is also shown in the figure also.  \\
\begin{table*}[]
\centering
\caption{\label{EXPERIMENTAL CROSS-SECTION} Measured K shell production cross-sections ({$\sigma_K^x$}) for Cu, Zn, and Ge targets with corresponding multiple ionization probabilities, modified fluorescence yields for multiple ionization \cite{lapicki2004effects} and ionization cross-sections ($\sigma_K^I$) by $^{28}$Si ions at different energies $(E)$ (MeV). Single vacancy fluorescence yields $(\omega_K^0)$ for Cu, Zn, and Ge are 0.454, 0.486, and 0.546, respectively \cite{krause1979atomic}, which very closed to \textcolor{black}{\citet{cipolla2011isics2011} values 0.44, 0.474, and 0.535, respectively
.} The cross-sections are in units of Kilobarn/atom.}
\begin{tabular}{|c|c|c|c|c|c|c|c|c|c|}
\hline
\multicolumn{10}{|l|}{\hspace{6.5cm}\textbf{Cu}}                                         \\ \hline
\textbf{E}   & \hspace{0.3cm}\textbf{$\sigma_K^x$} & \hspace{0.4cm}$\omega^0_K$& \hspace{0.3cm}$\sigma_K^I-SI$&P & $\omega_{K}$ & \hspace{0.3cm}$\sigma_K^I-MI$& ECUSAR& $\frac{ECUSAR}{\sigma_K^I-SI}$&$\frac{ECUSAR}{\sigma_K^I-MI}$ \\ \hline
84  & 27.4$\pm$1.4 &0.454&60.4$\pm$4.5 &0.817 & 0.816$\pm$0.03 & 33.5$\pm$2.2&17.1 &0.283 &  0.510     \\ \hline
90  & 34.2$\pm$1.8 & 0.454 & 75.3$\pm$5.6& 0.776 & 0.780$\pm$0.03 & 44.07$\pm$2.9& 21.6 & 0.287 & 0.491        \\ \hline
98  & 41.9$\pm$2.2& 0.454&92.3$\pm$6.8&0.736 & 0.738$\pm$0.03 & 56.89$\pm$3.8&  28.4& 0.308&  0.500        \\ \hline
107 & 50.5$\pm$2.6&0.454 &112.2$\pm$8.2& 0.649 & 0.701$\pm$0.03 & 77.8$\pm$4.8 &36.5 & 0.325&  0.469         \\ \hline
118 & 69.9$\pm$5.3&0.454 & 153.9$\pm$14.5&0.590 & 0.668$\pm$0.03 & 101.6$\pm$7.7 &  47.5& 0.301& 0.468    \\ \hline
128 & 89.3 $\pm$6.8&0.454 &196.7$\pm$18.5 & 0.547 & 0.645$\pm$0.03 & 138.4$\pm$10.2 & 58.0& 0.295& 0.419 \\ \hline
140 & 97.2$\pm$7.3&0.454 &214.1$\pm$20.0& 0.501 & 0.624$\pm$0.03 & 155.8$\pm$11.6  &70.9&0.331& 0.455\\ \hline
\multicolumn{10}{|l|}{\hspace{6.5cm}\textbf{Zn}}                                         \\ \hline
84  & 15.5$\pm$0.8&0.486 & 31.9$\pm$2.0&0.785 & 0.811$\pm$0.02 & 19.7$\pm$1.2& 11.6&0.364&0.590\\ \hline
90  & 17.9$\pm$0.9 &0.486 & 36.8$\pm$2.3&0.736 & 0.779$\pm$0.02 & 24.3$\pm$1.4  &  14.8&0.403& 0.610      \\ \hline
98  & 21.0$\pm$1.1&0.486 & 43.2$\pm$2.7&0.680 & 0.745$\pm$0.02 & 30.9$\pm$1.7   &19.7&0.456& 0.637   \\ \hline
107 & 25.5$\pm$1.3 &0.486 &52.5$\pm$3.3& 0.626 & 0.715$\pm$0.02 & 40.7$\pm$2.2   &25.8&0.492& 0.634       \\ \hline
118 & 46.9$\pm$3.7 &0.486 & 96.5$\pm$8.3& 0.571 & 0.686$\pm$0.02 & 68.4$\pm$5.7 &34.0 & 0.353& 0.498     \\ \hline
128 & 49.6$\pm$3.9 &0.486 &102.1$\pm$8.7& 0.529 & 0.666$\pm$0.02 & 74.5$\pm$6.2 &42.1 & 0.412& 0.565    \\ \hline
140 & 53.0$\pm$4.1  &0.486& 109.1$\pm$9.3& 0.486 & 0.647$\pm$0.02 & 81.9$\pm$6.8 &52.0 &0.477& 0.635      \\ \hline
\multicolumn{10}{|l|}{\hspace{6.5cm}\textbf{Ge}}                                         \\ \hline
84  & 9.4$\pm$0.5   &0.546 &17.2$\pm$1.0 & 0.724 & 0.809$\pm$0.02 & 11.6$\pm$710 &5.6&0.328&   0.486       \\ \hline
90  & 11.0$\pm$0.6  &0.546  & 20.1$\pm$1.2&0.679 & 0.787$\pm$0.02 & 13.9$\pm$0.8 &7.3& 0.363&  0.525     \\ \hline
98  & 13.6$\pm$0.7 &0.546 & 24.9$\pm$1.5&0.632 & 0.762$\pm$0.02 & 17.8$\pm$0.1 & 9.9 &0.397& 0.556        \\ \hline
107 & 16.7$\pm$0.9 &0.546 & 30.6$\pm$1.9&0.581 & 0.740$\pm$0.02 & 22.6$\pm$0.1 &13.3 &0.434&  0.587     \\ \hline
118 & 30.5$\pm$0.2 &0.546   &55.9$\pm$4.8& 0.534 & 0.717$\pm$0.02 & 42.5$\pm$0.3 &18.0 &0.322&0.423   \\ \hline
128 & 36.3$\pm$2.8 &0.546    &66.5$\pm$5.7& 0.495 & 0.703$\pm$0.02 & 51.6$\pm$3.6&22.7& 0.341& 0.440     \\ \hline
140 & 38.7$\pm$3.0   &0.546  & 70.9$\pm$6.0& 0.456 & 0.687$\pm$0.02 & 56.3$\pm$4.0 & 28.8 &0.406&  0.511      \\ \hline
\end{tabular}
\end{table*}
%
\begin{table*}[]
\centering
\caption{\label{AVERAGE PEAK ENERGY} Energy $E_x$ (keV) of $K_\alpha$ and $K_\beta$ x-rays of Cu, Zn and Ge targets for different beam energies (E) in MeV of $^{28}{Si}$ projectile. Their mean error of average energy of Cu, Zn, Ge for $K_\alpha$ and $K_\beta$ in keV are (8.01 $\pm$ 0.1, 9.04 $\pm$ 0.13), (8.55 $\pm$ 0.1, 9.72 $\pm$ 0.12) and (9.87 $\pm$ 0.1, 11.2 $\pm$ 0.1), respectively. $K_\alpha$ and $K_\beta$ energy of Cu, Zn and Ge targets in keV are (8.048, 8.905), (8.639, 9.572), and (9.886, 10.982), respectively.}
\begin{tabular}{|l|l|l|l|l|l|l|}
\hline
\multirow{2}{*}{\textbf{E}} & \multicolumn{2}{l|}{\hspace{1cm}\textbf{Cu}}         & \multicolumn{2}{l|}{\hspace{1cm}\textbf{Zn}}         & \multicolumn{2}{l|}{\hspace{1cm}\textbf{Ge}}         \\ \cline{2-7} 
                            & \textbf{\hspace{0.4cm}$K_\alpha$} & \textbf{\hspace{0.4cm}$K_\beta$} & \textbf{\hspace{0.4cm}$K_\alpha$} & \textbf{\hspace{0.4cm}$K_\beta$} & \textbf{\hspace{0.4cm}$K_\alpha$} & \textbf{\hspace{0.4cm}$K_\beta$} \\ \hline
84                          & 8.12$\pm$0.11       & 9.12$\pm$0.14      & 8.57$\pm$0.1        & 9.85$\pm$0.11      & 9.96$\pm$0.10       & 11.31$\pm$0.12     \\ \hline
90                          & 8.07$\pm$0.10       & 9.12$\pm$0.12      & 8.63$\pm$0.09       & 9.77$\pm$0.11      & 9.90$\pm$0.10       & 11.25$\pm$0.11     \\ \hline
98                          & 7.94$\pm$0.11       & 8.98$\pm$0.12      & 8.62$\pm$0.11       & 9.76$\pm$0.13      & 9.85$\pm$0.10       & 11.18$\pm$0.11     \\ \hline
107                         & 8.01$\pm$0.12       & 9.02$\pm$0.15      & 8.52$\pm$0.11       & 9.65$\pm$0.12      & 9.81$\pm$0.10       & 11.12$\pm$0.11     \\ \hline
118                         & 7.99$\pm$0.12       & 9.01$\pm$0.15      & 8.45$\pm$0.09       & 9.55$\pm$0.14      & 9.80$\pm$0.10       & 11.11$\pm$0.11     \\ \hline
128                         & 7.86$\pm$0.15       & 8.8$\pm$0.16       & 8.50$\pm$0.12       & 9.60$\pm$0.14      & 9.80$\pm$0.10       & 11.11$\pm$0.11     \\ \hline
140                         & 8.12$\pm$0.11       & 9.2$\pm$0.13       & 8.57$\pm$0.1        & 9.85$\pm$0.13      & 9.96$\pm$0.10       & 11.31$\pm$0.13     \\ \hline
\end{tabular}
\end{table*}
\indent The inner-shell vacancies are produced predominantly by the direct Coulomb ionization process, which can be treated perturbatively using the first-order perturbation approaches, namely, the plane-wave Born approximation (PWBA) \cite{choi1973tables}. The standard PWBA approach for direct ionization was further developed to include the hyperbolic trajectory of the projectile, the relativistic wave functions, and the corrections for the binding-polarization effect. The most advanced approach based on the PWBA, which goes beyond the first-order treatment to include the corrections for
the binding-polarization effects within the perturbed stationary states (PSS) approximation, the projectile energy loss
(E), and Coulomb deflection (C) effects as well as the relativistic (R) description of inner-shell electrons, is known as
the ECPSSR theory \cite{brandt1981energy}. This theory is further modified to replace the PSS effect by a united and separated atom (USA) treatment (ECUSAR) and valid in the complementary collision regimes of slow and intermediate to fast collisions, respectively \cite{lapicki2004effects}.\\
\begin{figure*}
\centering
\includegraphics[width=18.0cm,height=6.0cm]{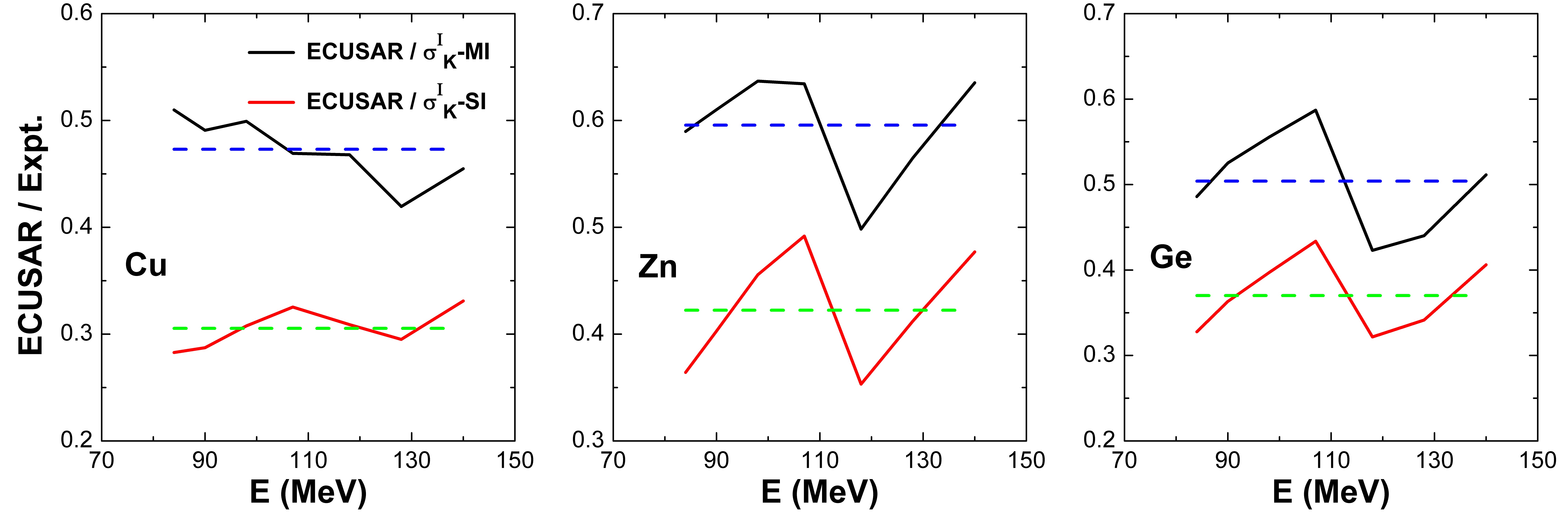}
\caption{Ratio of theoretical (ECUSAR) ionization cross-sections to the experimental ionization cross-sections is plotted as a function of beam energy. Experimental ionization cross sections are evaluated for single ionization in K-shell as well as for  single ionization in K-shell plus certain ionization in higher shells. Green and blue dashed lines indicate the average line for the ratio of theoretical ionization cross-sections to the experimental ionization cross-sections for single and multiple ionization, respectively.}.
\label{ratio theory vs expt.}
\end{figure*}
\indent In Fig. \ref{Expt and DI theories}, the measured $\sigma^I_{K}$ are plotted as a function of the beam energies, where the overall experimental uncertainty in the present cross-section measurements is attributed to the uncertainties in the photopeak, absolute efficiency of the detector, charge collected in Faraday cup, and target thickness. We have compared the measured $\sigma^I_{K}$ with the predictions of direct ionization cross-section from the theories ECPSSR and ECUSAR using the ISICS00 code \cite{bativc2012isicsoo}. We can see that the ECPSSR and ECUSAR predictions are almost equal. It just proves the fact that though $Z_1/Z_2$ for the systems used in the present experiment be in favour of united atom situation but at 3-5 MeV/u energy it ought to behave as the separated atom and thus the ECUSAR values are almost the same as the ECPSSR figures. Fig. \ref{ratio theory vs expt.} shows the ratio of theoretical (ECUSAR) ionization cross-sections to the experimental ionization cross-sections considering the fluorescence yield with single vacancy only in K-shell as well as with single vacancy in K-shell along with other vacancies in higher shells. In former case on the average the ratio is equal to 0.31, 0.42 and 0.37, respectively for Cu, Zn and Ge. Where as the ratios are increased to 0.48, 0.60 and 0.51, respectively for Cu, Zn and Ge.  \\
\indent As mentioned above, the measured $\sigma^I_{K}$ are much higher than the predictions. Further, we found that such findings are not at all observed for the first lime. In recent years a series of measurements by the Kazakhstan group \cite{gluchshenko2016argon,gorlachev2017k,gorlachev2018k} showed this fact very clearly in argon, Krypton and Xenon induced x-ray production cross-section measurements. The scenario becomes clearer when one reads the changes from light proton induced ionization \cite{batyrbekov2014k} to heavy xenon induced cases \cite{gorlachev2018k} through nitrogen \cite{batyrbekov2014x}, oxygen \cite{gorlachev2016oxygen}, neon \cite{gorlachev2019x}, argon \cite{gluchshenko2016argon} and krypton \cite{gorlachev2017k}. More importantly, the difference between the experiment and prediction is not only in the case of K-shell x-ray production cross sections but also in L-shell and M-shell x-ray production cross sections too.   \citet{gorlachev2018k} attributed the difference observed in the case of symmetric partners to the molecular mechanisms of the ion-atom interaction when mutual distortion of the atomic orbitals of the colliding partners can take place and the formation of a quasi molecule is expected. For the asymmetric and slow collisions, where $Z_1\gg Z_2$ and $v_1/v_{2S}$$\leq$1, the effect of electron capture of the target electrons to the vacant projectile shells becomes important. Further, accounting the multiple ionization of the target atoms by heavy ions is also important. All these effects may lead to a big change in the x-ray productions. Recently, \citet{masekane2020measurement} has attempted to take electron capture into account for the K-shell ionization. We have addressed this issue in a forthcoming article in a greater detail \cite{chatterjee2021investigation}. 
\section{Conclusion}
\indent We have measured K-shell ionization by heavy-ion impact. Here, the targets Cu, Zn, and Ge were bombarded by the 84-140 MeV $^{28}$Si ions to measure K-shell production cross-sections. We observed that the measured ionization cross-sections differ by at least a factor of two and three times higher than the theoretical direct ionization cross-sections if multiple ionization effects are taken and not taken into account. Electron capture from the target K-shell to the K- and L-shell of the projectile ions may be required to consider in resolving this difference. Appropriate theoretical work is extremely welcome to address this issue not only for fundamental understanding but also for practical applications in heavy-ion induced x-ray emission techniques for elemental analysis.

\section{acknowledgements}
\indent We acknowledge cooperation from the Pelletron accelerator staff during the experiments. S.S. gratefully acknowledges one of his supervisors B. R. Behera for his support throughout this work. \\

\bibliography{MANUSCRIPT.bbl}
\bibliographystyle{apsrev4-1}
\end{document}